\documentclass[useAMS,usegraphicx]{mn2e}
\usepackage[dvips]{graphicx}
\usepackage{longtable}
\usepackage{setspace}

\begin{document}

\title[Eccentricity of persistent sources]{Effect of pulse profile variations on measurement of eccentricity in orbits of Cen X-3 and SMC X-1}

\author[Raichur \& Paul]{Harsha Raichur$^{1,2}$, Biswajit Paul$^{1}$\\
1. Raman Research Institute, Sadashivanagar, C. V. Raman Avenue,
Bangalore 560\,080, India \\
2. IUCAA, Post bag 4, Ganeshkhind,
Pune 411\,007, India\\
(E-mail:harsha{@}iucaa.ernet.in,bpaul{@}rri.res.in)}
\date{Accepted.....; Received .....}
\maketitle

\begin{abstract}
It has long been argued that better timing precision allowed by satelites like Rossi X-ray Timing Experiments (RXTE) will allow us to measure the orbital eccentricity and the angle of periastron of some of the bright persistent high mass X-ray binaries (HMXBs) and hence a possible measurement of apsidal motion in these system. Measuring the rate of apsidal motion allows one to estimate the apsidal motion constant of the mass losing companion star and hence allows for the direct testing of the stellar structure models for these giant stars present in the HMXBs. In the present paper we use the archival RXTE data of two bright persistent sources, namely Cen X-3 and SMC X-1, to measure the very small orbital eccentricity and the angle of periastron. We find that the small variations in the pulse profiles of these sources rather than the intrinsic time resolution provided by RXTE, limit the accuracy with which we can measure arrival time of the pulses from these sources. This influences the accuracy with which one can measure the orbital parameters, especially the very small eccentricity and the angle of periastron in these sources. The observations of SMC X-1 in the year 2000 were taken during the high flux state of the source and we could determine the orbital eccentricity and $\omega$ using this data set.
\end{abstract}


\section{Introduction}
Formation of a high mass X-ray binary pulsar involves a supernova explosion giving birth to a neutron star. This initial supernova explosion imparts an impulse to the companion resulting in an eccentric binary orbit. However, most persistent high mass X-ray binaries observed have very low eccentricity. Tidal interaction between the neutron star and the high mass companion star are understood to circularize the binary orbit (Lecar, Wheeler, McKee 1976; Levine et al. 1993; and reference therein). There has been much interest in measuring the eccentricity and angle of periastron of the orbit of these persistent HMXBs using increased timing accuracy to determine the arrival time of X-ray pulses (Fabbiano \& Schreier 1977, Kelley et al. 1983, Primini, Rappaport \& Joss, 1977). Measuring the small eccentricity will allow one to constrain the age of the binary system since the supernova explosion and measuring the angle of pariastron at different epochs of time will allow to measure the rate of change of the angle of periastron. The rate of change of periastron angle (usually referred to as the rate of apsidal motion) is proportional to the apsidal motion constant of the companion star and hence can be used to test the standard stellar structure models for these massive O/B type stars residing in these binary systems. \\

In this paper we present results from analysis of the archival data from Rossi X-ray Timing Explorer (RXTE) observations of two persistent HMXBs Cen X-3 and SMC X-1. The pulsar nature of Cen X-3 was first discovered with \textit{Uhuru} (Scherier et. al. 1972; Giacconi et. al. 1971). It is an eclipsing binary having a spin period of $\sim 4.8$ s and an orbital period of $\sim 2.1$ days decaying at the rate of $(-1.78 \pm 0.08) \times 10^{-6} yr^{-1}$ (Kelley et. al. 1983). The binary system consists of a neutron star of mass $M_x = 1.21 \pm 0.21 M_{\odot}$ and a companion O6-8 III super giant star of mass $M_{opt} = 20.5 \pm 0.7 M_{\odot}$ (Hutchings et al. 1979; Ash et al. 1999). The distance to the binary system is estimated at 8 kpc, with a lower limit of 6.2 kpc (Krzeminski 1974).  \\

SMC X-1 also discovered using \textit{Uhuru} (Leong, et al. 1971) consists of a neutron star of spin period 0.71 s (Lucke et al. 1976) and a young B0 super giant companion (Webster et al. 1972, Liller 1973). There are clear X-ray eclipses which occur once every orbital period of 3.892 days (Schreier et al. 1972). The mass of the neutron star in this system is low $M_x = 1.06 \pm 0.1$ and the companion star is of mass $M_{opt} = 15.6 \pm 1.5$ (Val Baker, Norton \& Quaintrell 2005, van der Meer et. al. 2007). The upper limit to SMC X-1 eccentricity is 0.0007 (Primini, Rappaport and Joss, 1977). \\

In the following sections we describe our analysis and results for these two systems and discuss the implications of our results for further observations of these sources using future missions providing high timing accuracy X-ray measurements like ASTROSAT-LAXPC. 

\section{Observation and analysis}

\begin{figure} 
\centering
\includegraphics[height=80mm,angle=-90]{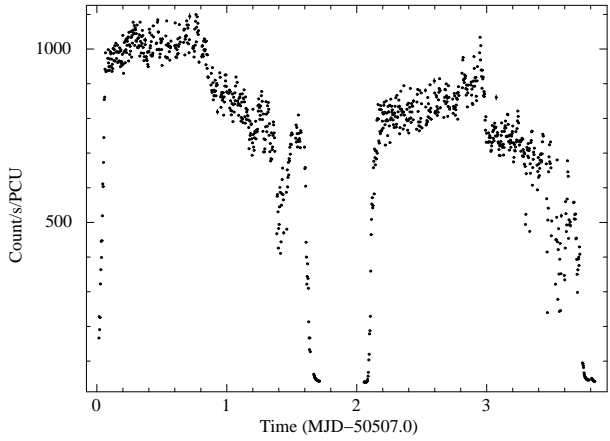}
\caption{Figure shows light curve of Cen X-3 for the 1997 observation made using RXTE-PCA. The light curve was generated using photons from only the first PCU}
\label{fig:cenx3-lc}
\end{figure}

\begin{figure} 
\centering
\includegraphics[height=80mm,angle=-90]{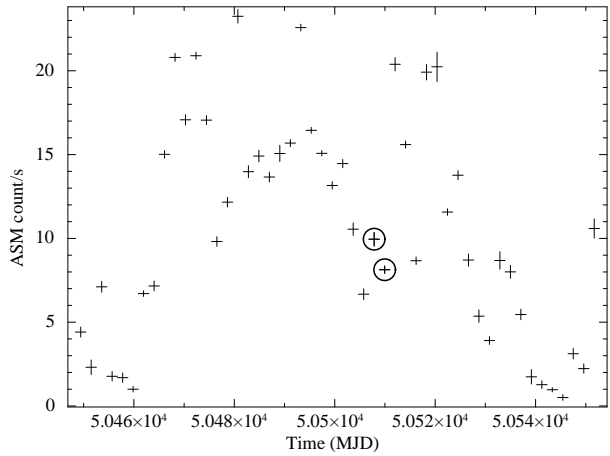}
\caption{Figure shows the ASM light curve of Cen X-3 during the 1997 observations. The points marked with circles show the time during which the PCA observations were carried out.}
\label{fig:cenx3-asm-lc}
\end{figure}

Cen X-3 was observed in 1997 for 3.8 days from February 28th to March 3rd using RXTE-PCA. Twelve pointed observations were made with five proportional counter units (PCU) for most of the time except for two occasions when only 4 PCUs were used. A total of 330 ks of data is available. Apart for the usual data modes Standard-I mode and Standard-II mode, data was also recorded in 2 single bit mode, 1 binned mode and 1 event mode data. The Standard-I mode has time resolution of 0.125s and no energy resolution whereas the Standard-II mode has 16s time resolution with the energy spectra divided into 129 channels binned according to a particular fixed scheme. For more information of RXTE-PCA data modes refer to the ABC of XTE guide available on the HEASARC site. The two single bit mode data together covered the energy range from 1 to 16 keV with 62 microsecond time resolution, the binned mode data also covered the same energy range with time resolution of 16 milliseconds and the event mode data covers the energy range above 16 keV with a time resolution of 31 microseconds. Figure \ref{fig:cenx3-lc} shows the complete light curve of this observation. From the long term All-Sky Monitor (ASM) light curve of Cen X-3 we can see that these observations were done when Cen X-3 was in an intermediate flux state (Figure \ref{fig:cenx3-asm-lc}). \\

\begin{figure} 
\centering
\includegraphics[height=80mm,angle=-90]{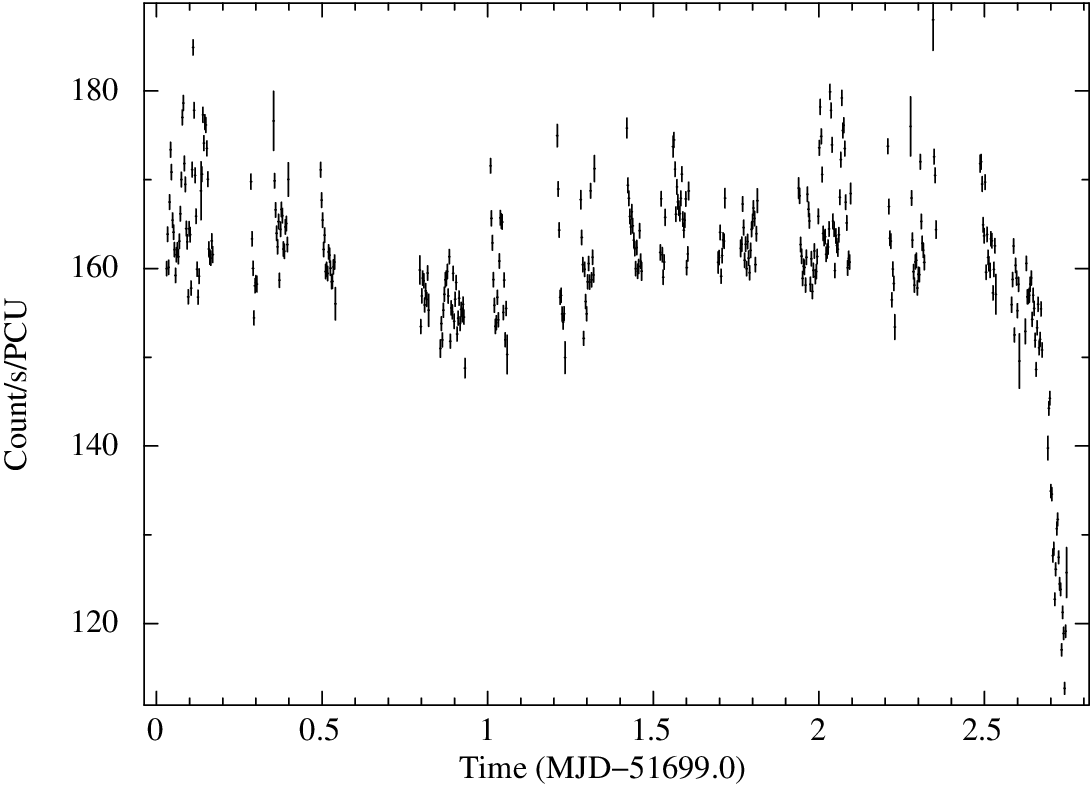}
\caption{Figure shows light curve of SMC X-1 made from the Standard-2 mode data of the 2000 observation.}
\label{fig:smcx1-2000-lc}
\end{figure}

\begin{figure} 
\centering
\includegraphics[height=80mm,angle=-90]{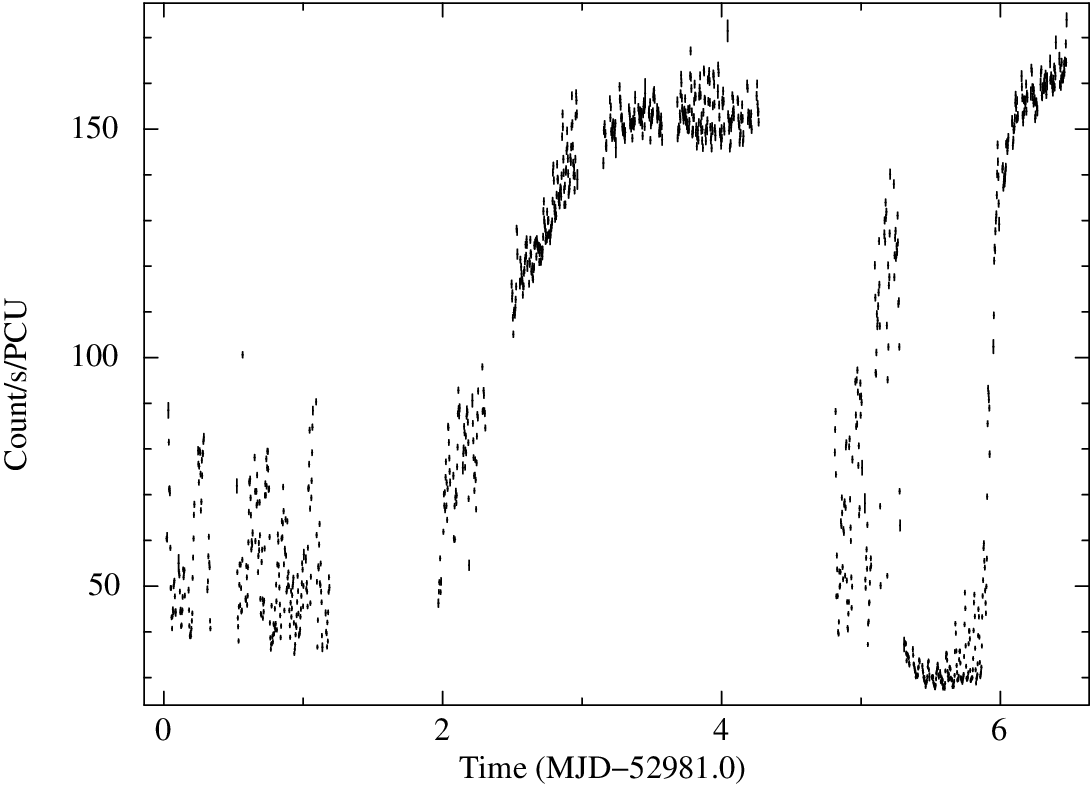}
\caption{Figure shows light curve of SMC X-1 made from the Standard-2 mode data of the 2003 observation.}
\label{fig:smcx1-2003-lc}
\end{figure}

\begin{figure} 
\centering
\includegraphics[height=80mm,angle=-90]{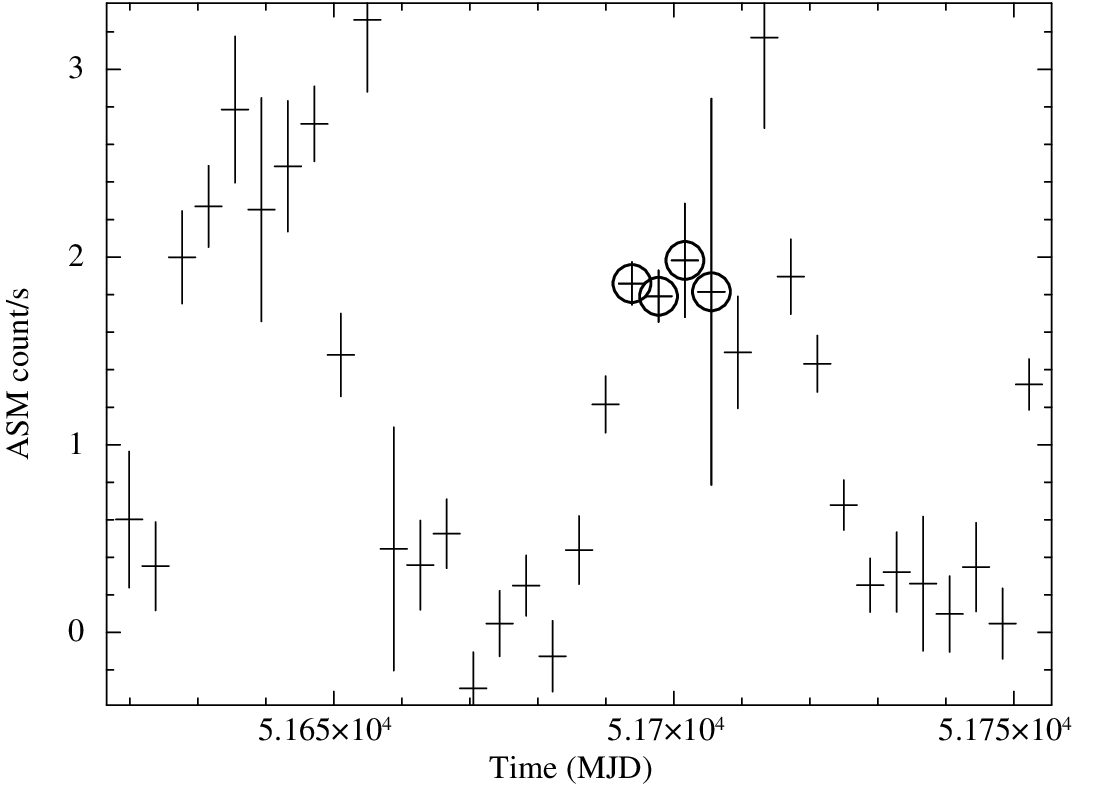}
\caption{Figure shows the ASM light curve of SMC X-1 during the 2000 observations. The points marked with circles show the time when the PCA observations were carried out.}
\label{fig:smcx1-asm-2000-lc}
\end{figure}

\begin{figure} 
\centering
\includegraphics[height=80mm,angle=-90]{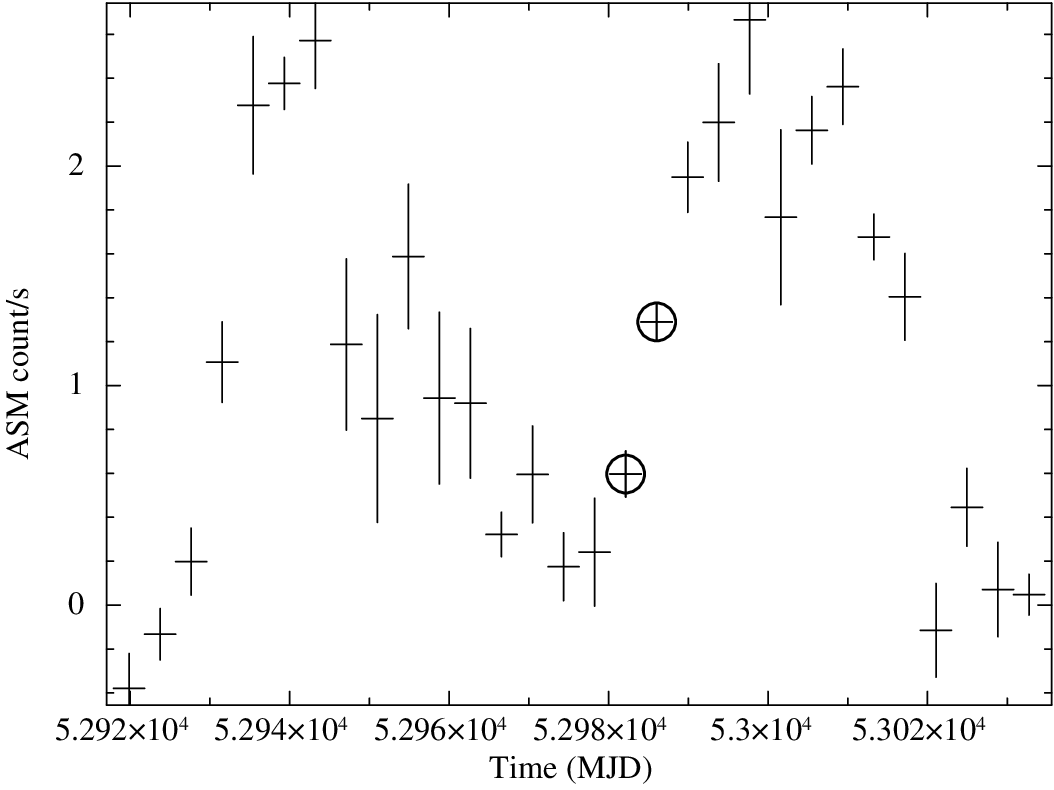}
\caption{Figure shows the ASM light curve of SMC X-1 during the 2003 observations. The points marked with circles show the time when the PCA observations were carried out.}
\label{fig:smcx1-asm-2003-lc}
\end{figure}

SMC X-1 was observed using RXTE in 2000 and 2003. It was observed for a total exposure of 234 ks in 2000 from MJD 51699 - MJD 51702 and for 557 ks in 2003 from MJD 52981 to MJD 52987. During both the observations, data was collected in Good Xenon mode, Standard I mode and the Standard II mode. Figure \ref{fig:smcx1-2000-lc} and Figure \ref{fig:smcx1-2003-lc} show the complete light curve of the observations for the year 2000 and 2003 respectively. SMC X-1 has a super orbital flux variation with quasi-periodicity of 50-60 days. The 2000 observation of SMC X-1 made using RXTE-PCA was done when the source was in its high state and the 2003 RXTE-PCA observation was done when the source was in a low state (see Figure \ref{fig:smcx1-asm-2000-lc} and \ref{fig:smcx1-asm-2003-lc}). 

\subsection{Pulse Profile variations} 
For absolute timing analysis the pulse profile of the neutron star should be stable. A changing pulse profile will introduce errors in measuring the arrival time of the pulses as one cannot have a unique time marker. So before presenting the details of the pulse arrival time analysis we first present the pulse profiles of both Cen X-3 and SMC X-1. For this we created a sample pulse profile from each pointed observation of both the sources. \\
 
\begin{figure} 
\centering
\includegraphics[height=80mm,angle=-90]{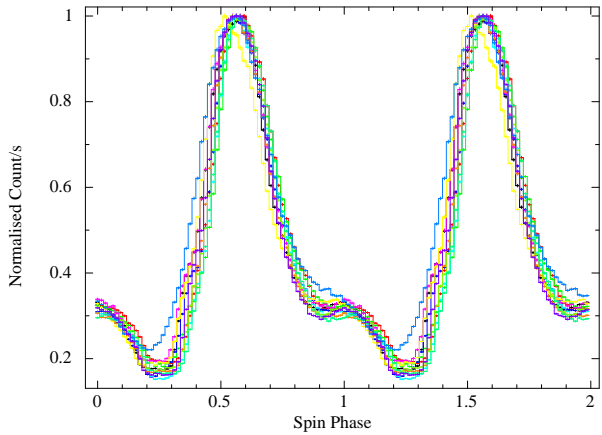}
\caption{Normalized sample pulse profiles of Cen X-3, each from the twelve pointed observations are 
shown with different colours. Since Cen X-3 count rate varies by factor of 2 during one orbital cycle the peak count rate per pulse profile is different. For comparison we have divided every pulse profile by its corresponding highest count rate.}
\label{fig:cenx3-pp}
\end{figure}

\begin{figure} 
\centering
\includegraphics[height=80mm,angle=-90]{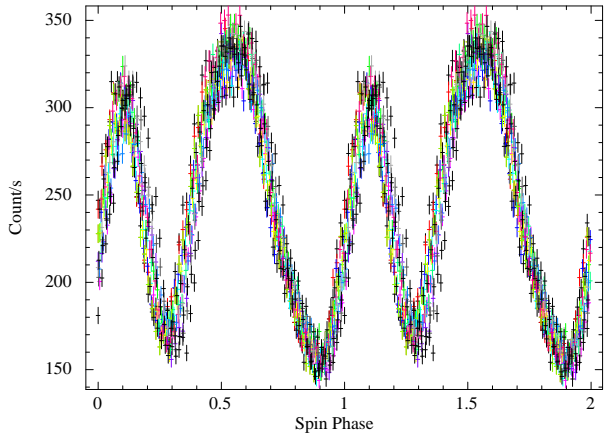}
\caption{Pulse profiles of SMC X-1 for all the 18 pointed observation of RXTE for the
year 2000 are shown. Each pulse profile is shown in a different colour. There are no drastic
variations in the pulse profiles. One hundred consecutive pulses were folded
together with the local spin period to generate each pulse profile.}
\label{fig:smcx1-pp}
\end{figure}

The pulse profile of Cen X-3 and SMC X-1 consists of two peaks, and we will refer to the highest peak as the primary peak and the other peak as the secondary peak. In the Cen X-3 pulse profile the secondary peak is much smaller than the primary peak whereas in the SMC X-1 pulse profile, both the peaks are of almost similar intensity. We generated a pulse profile for each pointed observation. The light curves used to generate the pulse profiles have time resolution of 15.625 ms and the photon arrival times were corrected to the solar system barycenter. Each light curve (one for every pointed observation) was searched for the local spin period ($P_{spin}$). The pulse profiles were then generated by folding the light curves with the corresponding $P_{spin}$. One hundred consecutive pulses were averaged to generate each pulse profile. Figure \ref{fig:cenx3-pp} and \ref{fig:smcx1-pp} show the pulse profiles of each pointed observation of Cen X-3 and SMC X-1 (2000 observation only) respectively. \\ 

Excluding the eclipse times, the instantaneous flux of Cen X-3 varies by a factor of 2 during one orbital cycle. As the neutron star is obscured by the companion star we receive more scattered X-rays during the eclipse, eclipse ingress and eclipse egress, due to which the pulse fraction varies (Paul, Raichur \& Mukherjee 2005; Raichur \& Paul 2008). The average shape of the pulse profiles look similar except for the peak count rates. To generate Fig \ref{fig:cenx3-pp} we have divided the count rates of each pulse profile by its respective peak count rate. This normalisation was adopted to bring out the small variations in the pulse shape even after subtracting the effect of varying instantaneous X-ray flux of the source. \\

In case of SMC X-1, the out of eclipse count rates do not vary drastically with orbital phase at least for the observations done during the year 2000. Thus no normalisation was used to generate Figure \ref{fig:smcx1-pp}. The pulse profiles do not show sharp variations at different orbital phases. \\

\begin{figure} 
\centering
\includegraphics[height=80mm,angle=-90]{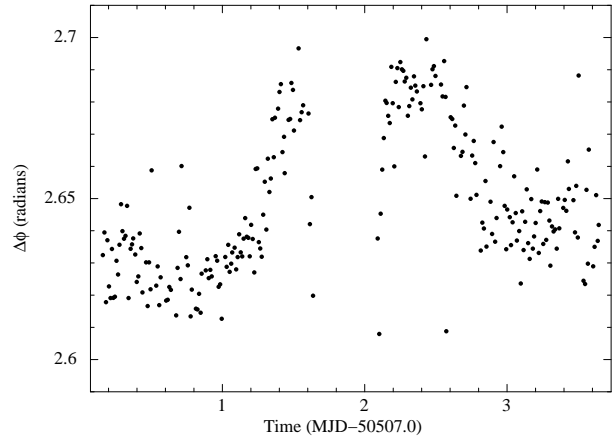}
\caption{This plot shows variation in $\Delta \phi$ of Cen X-3 pulse profile as a function of observation time. 
The structure seen in the plot indicates that the spin phase of primary peak and secondary peak are not locked and hence can introduce errors in the measurement of pulse arrival times.} 
\label{fig:cenx3-delta-phi}
\end{figure}

\begin{figure} 
\centering
\includegraphics[height=80mm,angle=-90]{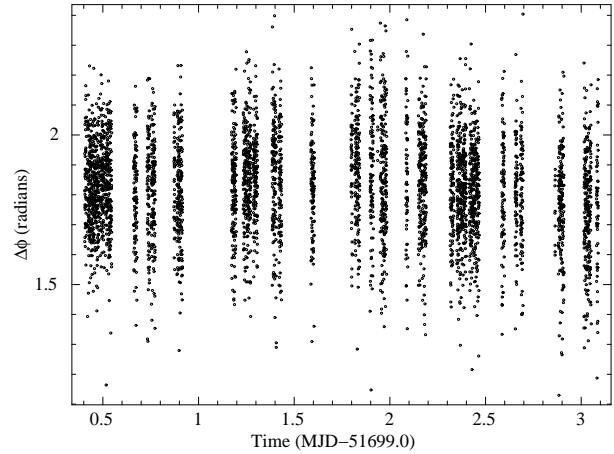}
\caption{Plot shows variations in $\Delta \phi$ of SMC X-1 as a function of observation time for the 2000 observations. No systematic structure is
seen. All data points that lie in the orbital phase of X-ray eclipse have been removed.}
\label{fig:smcx1-delta-phi-2000}
\end{figure}

\begin{figure} 
\centering
\includegraphics[height=80mm,angle=-90]{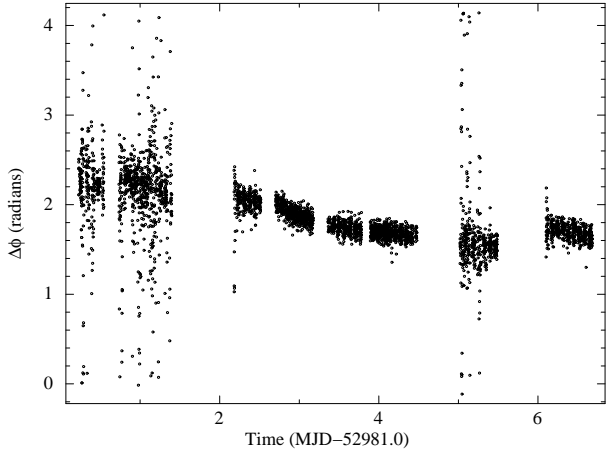}
\caption{$\Delta \phi$ of SMC X-1 is plotted as a function of observation time. The 2003 observations show larger values of 
$\Delta \phi$ as compared to those of the 2000 observations and there are also some systematic variations. All data points that lie in the orbital phase of X-ray eclipse have been removed.}
\label{fig:smcx1-delta-phi-2003}
\end{figure}

To look for systematic variations in the pulse profiles of both Cen X-3 and SMC X-1 we Frourier transform the pulse profiles and calculated $\Delta \phi$, a quantity which we define as the difference between the spin phase of the first and second highest Fourier component ($\phi_1$ and $\phi_2$ respectively).
$\Delta \phi$ is related to the difference in the spin phase of the primary peak and secondary peak of the pulse profile. Figure \ref{fig:cenx3-delta-phi}, \ref{fig:smcx1-delta-phi-2000} and \ref{fig:smcx1-delta-phi-2003} show the variations in $\Delta \phi$ as a function of orbital phase for Cen X-3 1997 observation and SMC X-1 2000 and 2003 observations respectively.  \\

\begin{figure} 
\centering
\includegraphics[height=80mm,angle=-90]{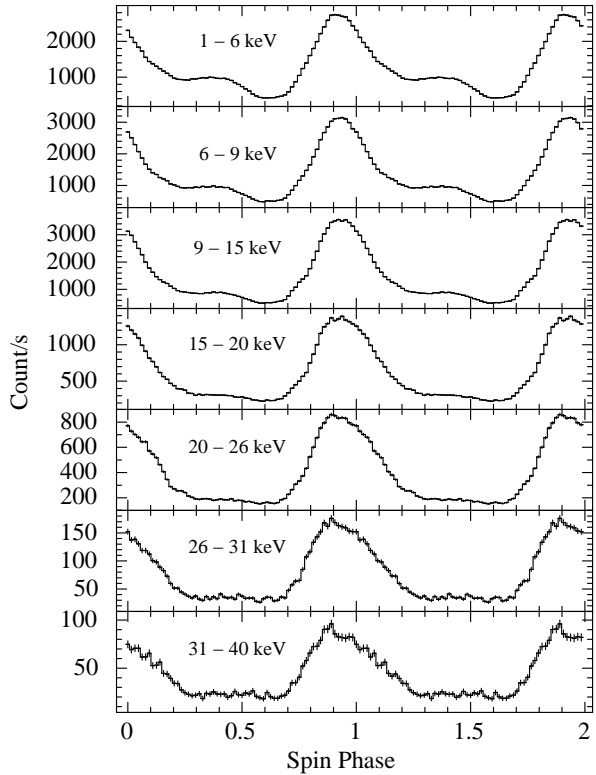}
\caption{Pulse profiles of Cen X-3 at different energy ranges folded with the local spin period and averaged over 100 consecutive pulses are shown. The primary peak remains almost similar at all energies but the secondary peak vanishes at higher energies.}
\label{fig:cenx3-energyresolved-pp}
\end{figure}

\begin{figure} 
\centering
\includegraphics[height=80mm,angle=-90]{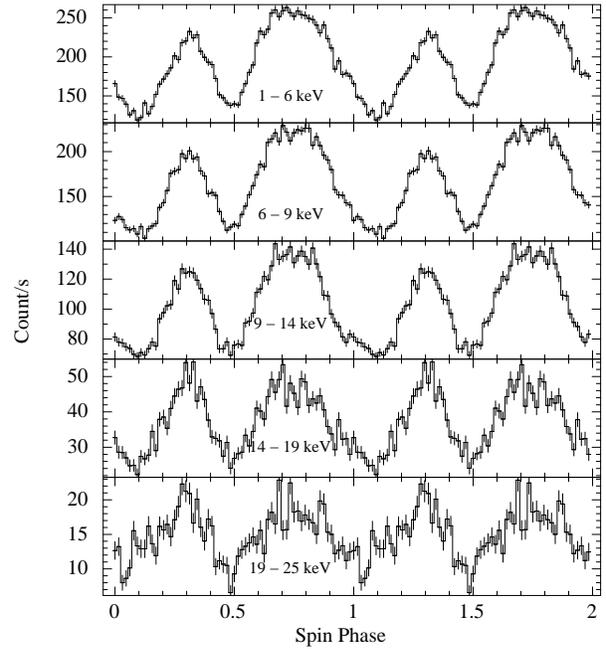}
\caption{Pulse profiles of SMC X-1 in different energy bands are shown. At lower energies the second peak of the pulse profile is shorter in amplitude compared to the first peak. At higher energies both the peaks in the pulse profile have similar amplitude. The 2000 observation data were used to get these energy resolved pulse profiles.}
\label{fig:smcx1-energyresolved-pp}
\end{figure}

 We also generated energy resolved pulse profiles for both Cen X-3 and SMC X-1. Figure \ref{fig:cenx3-energyresolved-pp} and \ref{fig:smcx1-energyresolved-pp} show the energy resolved pulse profiles of Cen X-3 and SMC X-1 respectively. In the pulse profiles of Cen X-3, we see that at higher energy, the secondary peak vanishes but the primary peak remains the same. In SMC X-1, at higher energies, both the primary peak and the secondary peak have similar amplitudes whereas at lower energies the secondary peak is smaller that the primary peak.

\subsection{Pulse arrival time analysis and estimation of orbital parameters:}
The arrival time of pulses from a neutron star in binary orbit are modified due to its motion in the orbit. When the neutron star is moving towards the observer the pulses arrive faster and when it is moving away from the observer pulses are delayed. These variations in arrival times of pulses are used to determine orbit of the neutron star. The emission ($t^{\prime} _n$) and arrival time of pulses ($t_n$) are related via the neutron star orbit ($f_{orb}(t^{\prime} _n)$) as  \\
\begin{eqnarray}
t^{\prime}_n & = & t_0 + nP_{spin} + \frac{1}{2} n^2 \dot P_{spin} P_{spin} \nonumber \\
t_n & = & t^{\prime}_n + f_{orb}(t^{\prime}_n)
\label{eqn:general-orbit-eqn}
\end{eqnarray}
For a nearly circular orbit, $f_{orb}$ is given by
\begin{eqnarray}
f_{orb} & = & a_x \sin i \cos l_n \nonumber \\
l_n & = & 2 \pi (t^{\prime}_n - E)/ P_{orb} + \pi /2
\label{eqn:circular-orbit}
\end{eqnarray}
where $l_n$ is a mean orbital longitude at time $t^{\prime} _n$ and is defined such that for an orbit with finite eccentricity and $\omega$, \textit{l} is related to mean anomaly M, by $l = M + \omega$. This definition of $l$ ensures that the mean and true longitudes agree at the apsides and it also remains well defined as the eccentricity vanishes, which is not true for mean anomaly. After measuring the arrival time of pulses, eqn \ref{eqn:circular-orbit} can be solved using the method of least squares and get initial estimates for $a_x \sin i$ and $E$. Note that $P_{orb}$ is better estimated by combining the history of orbital epochs $E_n$'s, where the nth orbital epoch is given by $E_n = E_0 + n P_{orb} + \frac{1}{2} n^2 P_{orb} \dot P_{orb}$. The residuals to the fit of observed arrival time delay of pulses are computed by subtracting the arrival times calculated using the initial estimates from the observed arrival times. The residual will show variations if the orbit has a small measurable non-zero eccentricity and/or if the initial estimates are not very accurate. In such a case the residuals can be corrected for by introducing the terms initially dropped from $f_{orb}$ namely $e$ and $\omega$ and the terms for differential corrections to the initial estimates of $a_x \sin i $, $E$. Such a function is given by (Luyten 1936; Sterne 1941; Russell 1902 and references therein)
\begin{eqnarray}
 \delta t_n & = & \delta t_0 + n \delta P_0 + \frac{1}{2} n^2 P_0 \dot P_0 \nonumber \\
            &   & + ~ \delta x \sin l_n - \frac{2 \pi x}{P_{orb}}\delta E \cos l_n \nonumber \\
            &   & + ~\frac{3}{2} x e \sin \omega + \frac{1}{2} x e \cos \omega \sin 2 l_n \nonumber \\
            &   & - ~\frac{1}{2} x e \sin \omega \cos 2 l_n
\label{eqn:residue-eqn}
\end{eqnarray}
If we substitute $g = e \sin \omega$ and $h = e \cos \omega$ in the above equation it becomes fully linear in the differential corrections and small parameters and can be solved using the method of linear least-squares.

\begin{figure} 
\centering
\includegraphics[height=80mm,angle=-90]{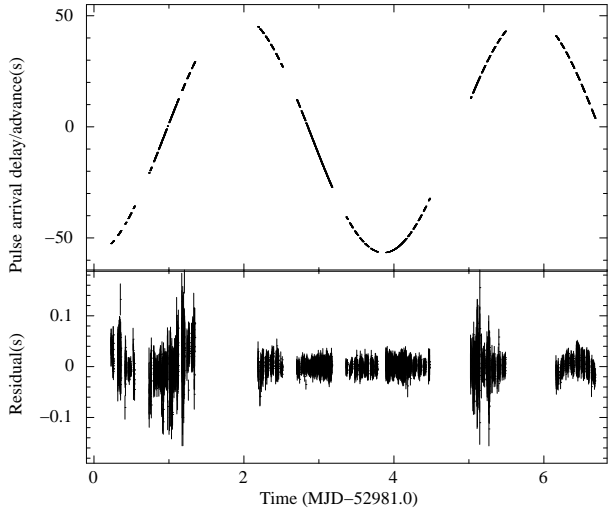}
\caption{Upper panel shows the delay curve of SMC X-1 obtained using the 2003 RXTE-PCA observations. Lower panel shows the residual pulse arrival delays/advance after subtracting delays expected by a circular orbit from the observed delays.}
\label{fig:arrival-time-delay-smcx1-2003}
\label{fig:smcx1-residues-2003}
\end{figure}

To determine the arrival times of pulses we followed the following procedure. For Cen X-3 we used the binned mode data to extract a light curve of time resolution 15.625 ms and covering an energy range of 2-35 keV. For SMC X-1 light curves were generated using good Xenon data with the same time resolution and using the full available energy coverage. The photon arrival times were corrected to the solar system barycenter. Individual pulse profiles at different orbital phases are generated by averaging a few consecutive pulse profiles. For Cen X-3 ten consecutive pulses and for SMC X-1 twenty consecutive pulses were averaged. The pulse profiles were Fourier transformed and the phase of the frequency component with the highest amplitude was used as the time marker. The arrival time delay of that pulse was then calculated using the product of the phase of highest frequency component and spin of the neutron star. Fig \ref{fig:arrival-time-delay-cenx3} is the delay curve of Cen X-3 and Fig \ref{fig:arrival-time-delay-smcx1-2000} and Fig \ref{fig:arrival-time-delay-smcx1-2003} are delay curves for SMC X-1 for the 2000 and 2003 respectively. \\

The delay curve is used to solve eqn \ref{eqn:general-orbit-eqn} for a circular orbit as given in eqn \ref{eqn:circular-orbit}. The projected semi-major axis and epoch E were free parameters. The residual delay and advance in the arrival time of pulses are calculated by subtracting the calculated delays from the observed delays. From the respective residual curves for the sources, we can see that only the residuals of SMC X-1 for the year 2000 have a sinusoidal variation. This residual curve of SMC X-1 was used to get the eccentricity and angle of periastron of the neutron star orbit. The residual curve of Cen X-3 and SMC X-1 (of the year 2003) have variations coupled due to a very small eccentricity and the variations in the pulse shape during the observations. These residual curves hence could not be used to get the eccentricity or $\omega$ of the orbit. \\

\section{Results and Discussions}
\subsection{Cen X-3:}
\begin{figure} 
\centering
\includegraphics[height=80mm,angle=-90]{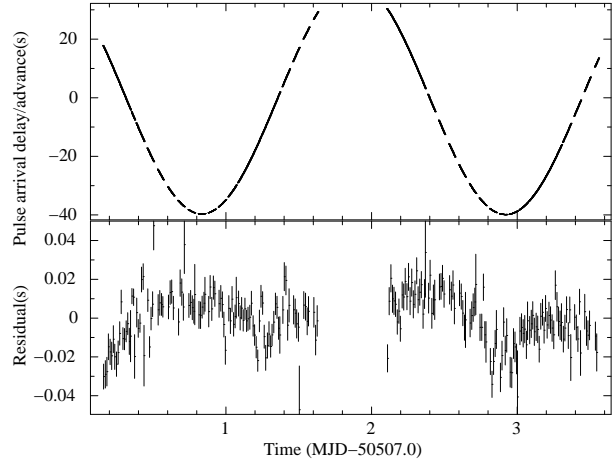}
\caption{In the upper panel of the figure are plotted arrival time delay/advance of pulses of Cen X-3 as function of time. The two orbits are clearly seen separated by an eclipse when no delay measurements are possible. This curve is used to solve Equation \ref{eqn:circular-orbit} and hence measure the orbital elements. In the lower panel are plotted the residual arrival time delay/advance of pulses as function of time. The residuals are obtained by subtracting the model arrival times expected due to a circular orbit from the observed pulse arrival delay curve.}
\label{fig:arrival-time-delay-cenx3}
\label{fig:cenx3-residues}
\end{figure}

\begin{figure} 
\centering
\includegraphics[height=80mm,angle=-90]{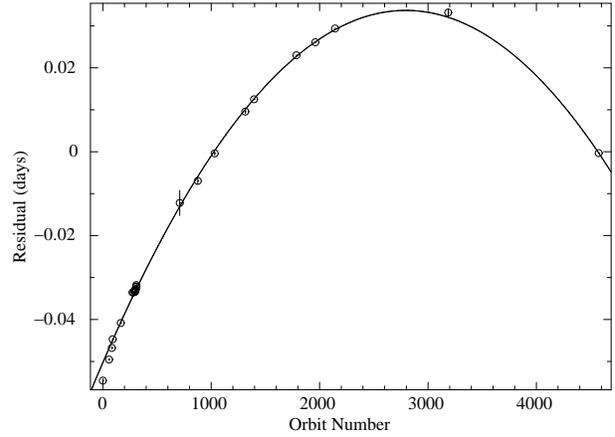}
\caption{The observed-minus-calculated eclipse times are plotted with respect to
the orbit number of Cen X-3. The latest data point is from this work. The quadratic trend due to orbital period decay is clearly 
seen. The solid curve represents the best-fit to a constant rate of orbital 
decay.}
\label{fig:cenx3-orbit-decay}
\end{figure}

The delay curve of Cen X-3 (Figure \ref{fig:arrival-time-delay-cenx3}) is used to solve equation \ref{eqn:general-orbit-eqn} with $a_x \sin i$ and $E$ as free parameters. To find the accurate $P_{orb}$ we combined the new value of $E$ with the previous measurements made by other authors (see Table \ref{tab:cenx3-E}). Figure \ref{fig:cenx3-orbit-decay} shows the observed-minus-calculated eclipse times plotted with respect to the orbit number of Cen X-3. The quadratic trend due to orbital period decay is clearly seen. Solid curve in the figure represents the best-fit to a constant rate of orbital decay.

Figure \ref{fig:cenx3-residues} shows the residual arrival time delay of the pulses. To clearly bring up the structure in the residuals we have averaged fifteen consecutive data points to get one point of the plot. A residual only due to a small eccentricity of the orbit would have been sinusoidal with a period half the orbital period of the neutron star. The amplitude of such a residual would be $\frac{1}{2}e a_x \sin i$ (Fabbiano \& Schierer, 1977). But the structure seen in the residuals is not purely sinusoidal. The non-sinusoidal component of the residuals could be due to the varying pulse profile of Cen X-3. As seen from the Figure \ref{fig:cenx3-delta-phi} the phase difference between the first highest Fourier component and the second highest Fourier component is not constant but is dependent on the orbital phase. From the energy resolved pulse profiles (see Figure \ref{fig:cenx3-energyresolved-pp}) it can be noted that the soft X-ray photons contribute more to the secondary peak. At orbital phases close to the eclipse, the X-ray photons reaching the observer have to pass through higher absorption column density, which affects the soft photons more than the hard photons. Also it has been noted that the pulse fraction of Cen X-3 is dependent on the average flux of the source (Raichur \& Paul 2008). The 1997 observations used for this work have been carried out during the intermediate flux state of Cen X-3 (for a more detailed definition of source flux state depending on the ASM count rate see Raichur \& Paul, 2008). Therefore the structure seen in the delay residuals is both due to a small eccentricity and a variable pulse profile, which limits the accuracy with which we can measure the arrival time of the pulses. We therefore conclude that observations done during low or intermediate flux state of Cen X-3 cannot be successfully used to determine the possible very small orbital eccentricity of the orbit. The amplitude of a sine curve fit to the residuals is $2.84 \times 10^{-3}$ lt-sec. Note that the sine curve has a fixed period of $P_{orb}/2$. Equating this amplitude to $\frac{1}{2}e a_x \sin i$ we get an upper limit for eccentricity of $0.0001$. The other orbital parameters and the revised rate of orbital decay of the system are tabulated in \ref{tab:cenx3-orbit-parameters}.

\subsection{SMC X-1}

\begin{figure} 
\centering
\includegraphics[height=80mm,angle=-90]{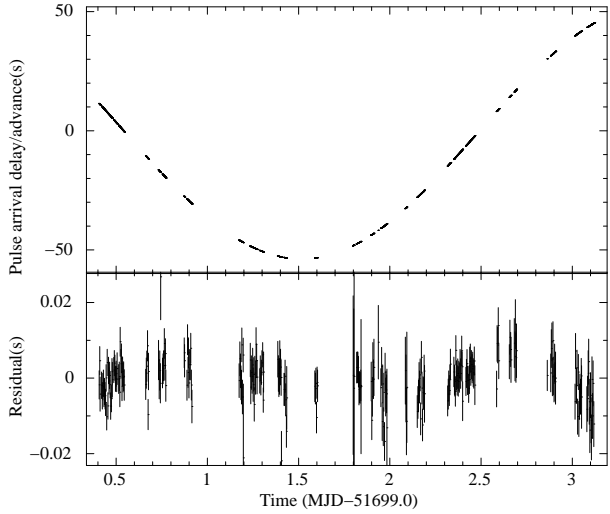}
\caption{The pulse arrival time delay curve for SMC X-1 obtained by analysis of the available 2000 observations is shown in the upper panel. The residuals obtained by subtracting the model arrival times expected due to a circular orbit from the observed pulse arrival delay curve are shown in the lower panel. For clarity 20 consecutive data points have been averaged to get one data point of the residual curve.}
\label{fig:arrival-time-delay-smcx1-2000}
\label{fig:smcx1-residues-2000}
\end{figure}

The delay curves of SMC X-1 for the 2000 and 2003 observations (Figure \ref{fig:arrival-time-delay-smcx1-2000} and \ref{fig:arrival-time-delay-smcx1-2003} respectively) are used to solve Equation \ref{eqn:general-orbit-eqn}. Similar to the Cen X-3 solution, $a_x \sin i$ and $E$ are the free parameters. The values of $a_x \sin i$ and $E$ for the two fits are given in Table \ref{tab:smcx1-orbit-parameters}. As already mentioned the 2000 observation was taken when the source was in high flux state and the 2003 observations when the source was in low flux state. The pulse fraction of SMC X-1 is a function of the average source flux, higher pulse fraction at higher flux levels (Kaur et. al. 2007). Also during the 2003 observations the source seems to be flaring, because of which the pulse profile is changing more rapidly. This has affected the accuracy with which the 2003 pulse timing analysis can be made. From the energy resolved pulse profiles, it can be seen that at soft energies the primary peak count rate is higher than the secondary peak count rate. But at harder energies both the secondary and primary peaks have similar count rate. Hence for the 2003 observations, the pulse profile has an orbital phase as well as average source flux dependence. Therefore the 2003 observations could not be used for measuring the small eccentricity of the NS orbit whereas the arrival time delay residuals of 2000 observation show only a sinusoidal variation which is not due to pulse profile evolution (see Figure \ref{fig:smcx1-residues-2000}). Therefore these residuals are fit using equation \ref{eqn:residue-eqn} which has $e$ and $\omega$ as free parameters along with corrections for $a_x \sin i$ and $E$. The estimated value of $e$ and $\omega$ are given in table \ref{tab:smcx1-orbit-parameters}.

\begin{figure} 
\centering 
\includegraphics[height=80mm,angle=-90]{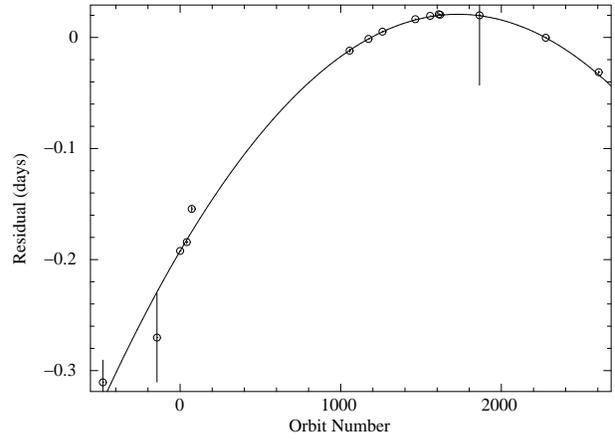} 
\caption{The observed-minus-calculated eclipse times for SMC X-1 are plotted 
with respect to the orbit number. Last two data points are from this work. The quadratic trend due to orbital period 
decay is clearly seen. The solid curve represents the best-fit to a constant 
rate of orbital decay.} 
\label{fig:smcx1-orbit-decay}
\end{figure}

The orbital period $P_{orb}$ and the rate of change of orbital period $\dot P_{orb}$ are revised by combining the two new orbital ephemeris with the previously known orbital ephemeris. Table \ref{tab:smcx1-E} gives the full list of previously determined orbital epochs. Figure \ref{fig:smcx1-orbit-decay} shows the observed-minus-calculated eclipse times plotted with respect to the orbit number of SMC X-1. The quadratic trend due to orbital period decay is clearly seen. 

From the above pulse analysis study we have one very striking conclusion that for even the brightest persistent XBPs evolution of pulse profiles must not be neglected. If the XBPs have super-orbital variations like in the case of both SMC X-1 and Cen X-3, then care must be taken to first establish if the pulse profile is average flux dependent. If there is dependency of the pulse profile on average source flux, then the error in timing due to such a dependency cannot be neglected for the present day high timing accuracy measurements possible with RXTE-PCA and future missions like ASTROSAT. For any work involving high absolute timing accuracy of an XBP which has super-orbital flux variations and has pulse fraction and pulse profile dependency on the average flux state, care must be taken that the observations are taken at such a time that the overall error due to these factors is minimal.

\section{Acknowledgements}
This research has made use of data obtained from the High Energy Astrophysics Science Archive Research Center (HEASARC), provided by NASA's Goddard Space Flight Center.

{}

\clearpage

\begin{table}
\centering
\caption{Cen X-3 orbit parameters measured using 1997 RXTE observations}
\begin{tabular}{|c|c|}
\hline
Parameter&Value \\
\hline
$a_x \sin i$&$39.6612 \pm 0.0009$ (lt-sec)\\
$E$&$50506.788423 \pm 0.000007$ (MJD)\\
$E_0$&$958.550276 \pm 0.0006$ (MJD)\\
$P_{orb}$&$2.08713936 \pm 0.00000007$ d\\
$\dot P_{orb}/P_{orb}$& $-(1.799 \pm 0.002) \times 10^{-6}$ yr$^{-1}$\\
\hline
\end{tabular}
\label{tab:cenx3-orbit-parameters}
\end{table}

\begin{table}
\begin{minipage}{160mm}
\centering
\caption{Cen X-3 orbital epoch history}
\begin{tabular}{|ccc|}
\hline
 & Time & \\
Orbital Cycle&JD - 2,440,000.0&Reference \\
\hline
0&$958.84643 \pm 0.00045$&Fabbiano \& Schreier 1977\\
57&$ 1077.81497 \pm 0.00015$&Fabbiano \& Schreier 1977\\
83&$ 1132.08181 \pm 0.00029$&Fabbiano \& Schreier 1977\\
91&$ 1148.78051 \pm 0.00016$&Fabbiano \& Schreier 1977\\
166&$ 1305.31533 \pm 0.00014$&Fabbiano \& Schreier 1977\\
273&$ 1528.64010 \pm 0.00030$&Fabbiano \& Schreier 1977\\
284&$ 1551.59798 \pm 0.00017$&Fabbiano \& Schreier 1977\\
293&$ 1570.38199 \pm 0.00011$&Fabbiano \& Schreier 1977\\
295&$ 1574.55610 \pm 0.00013$&Fabbiano \& Schreier 1977\\
296&$ 1576.64330 \pm 0.00010$&Fabbiano \& Schreier 1977\\
297&$ 1578.73037 \pm 0.00007$&Fabbiano \& Schreier 1977\\
298&$ 1580.81722 \pm 0.00009$&Fabbiano \& Schreier 1977\\
300&$ 1584.99193 \pm 0.00010$&Fabbiano \& Schreier 1977\\
303&$ 1591.25328 \pm 0.00015$&Fabbiano \& Schreier 1977\\
304&$ 1593.34025 \pm 0.00015$&Fabbiano \& Schreier 1977\\
307&$ 1599.60212 \pm 0.00015$&Fabbiano \& Schreier 1977\\
308&$ 1601.68930 \pm 0.00014$&Fabbiano \& Schreier 1977\\
309&$ 1603.77671 \pm 0.00021$&Fabbiano \& Schreier 1977\\
709&$ 2438.628 \pm 0.003$&Tuohy 1976\\
876&$ 2787.1755 \pm 0.0007$&van der Klis, Bonnet-Bidaud, \& Robba 1980\\
1032&$ 3112.76642 \pm 0.0004$&Kelly et al. 1983\\
1314&$ 3701.33275 \pm 0.00043$&Howe et al. 1982\\
1395&$ 3870.38910 \pm 0.00002$&Kelly et al. 1983\\
1786&$ 4686.44760 \pm 0.00005$&Murakami et al. 1983\\
1960&$ 5049.6025 \pm 0.0001$&Nagase et al. 1984\\
2142&$ 5429.45421 \pm 0.00005$&Nagase et al. 1984\\
3186&$ 7608.3688 \pm 0.0008$&Nagase et al. 1992\\
4575&$ 10507.288423 \pm 0.000007$&present work\\
\hline
\end{tabular}
\label{tab:cenx3-E}
\end{minipage}
\end{table}

\clearpage

\begin{table}
\begin{minipage}{160mm}
\centering
\caption{SMC X-1 orbit parameters}
\begin{tabular}{||c|c c||}
\hline
\hline
Parameter&\multicolumn{2}{c|}{Value} \\
& 2000 & 2003 \\
\hline
$a_x \sin i$&$53.5769 \pm 0.0006$ (lt-sec)& $53.339 \pm 0.032$ (lt-sec)\\
$E$&$51694.673697 \pm 0.000012$ (MJD)&$52979.0159 \pm 0.0022$ (MJD)\\
$e$&$0.00021 \pm 0.00001$& -- \\
$\omega$&$-43^0.23 \pm 8^0.99 $& -- \\
$E_0$&\multicolumn{2}{c||}{$42836.1827 \pm 0.0003$ (MJD)} \\
$P_{orb}$&\multicolumn{2}{c||}{$3.89229263 \pm 0.0000004 $ d } \\
$\dot P_{orb}/P_{orb}$&\multicolumn{2}{c||}{$-(3.414 \pm 0.003) \times 10^{-6}$ yr$^{-1}$} \\
\hline
\hline
\end{tabular}
\label{tab:smcx1-orbit-parameters}
\end{minipage}
\end{table}


\begin{table}
\begin{minipage}{160mm}
\centering
\caption{SMC X-1 orbital epoch history}
\begin{tabular}{|ccc|}
\hline
 & Time & \\
Orbital Cycle&MJD&Reference \\
\hline
-481&$ 40963.99 \pm 0.02$&Scherier et al. 1972\\
-144&$ 42275.65 \pm 0.04$&Tuohy \& Rapley 1975\\
0   &$ 42836.1828 \pm 0.0002$&Primini et al. 1977\\
42  &$ 42999.6567 \pm 0.0016$&Davison 1977\\
72  &$ 43116.448  \pm 0.0022$&Bonnet-Bidaud et al. 1981\\
1055&$ 46942.47237 \pm 0.0015$&Levine et al. 1993\\
1173&$ 47401.744476 \pm 0.000007$&Levine et al. 1993\\
1260&$ 47740.35906  \pm 0.00003$&Levine et al. 1993\\
1464&$ 48534.34786  \pm 0.00035$&Wojdowski et al. 1998\\
1556&$ 48892.4191 \pm   0$&Wojdowski et al. 1998\\
1610&$ 49102.59109 \pm  0.00082$&Wojdowski et al. 1998\\
1619&$ 49137.61911 \pm 0.00050$&Wojdowski et al. 1998\\
1864&$ 50091.170  \pm  0.063$&Wojdowski et al. 1998\\
2276&$ 51694.673022\pm 0.00001$&Present work\\
2606&$ 52979.0174 \pm  0.001$&Present work\\
\hline
\end{tabular}
\label{tab:smcx1-E}
\end{minipage}
\end{table}

\clearpage

\end{document}